*Paper*

# Planetary nebula morphologies indicate a jet-driven explosion of SN 1987A and other core-collapse supernovae

Noam Soker[1]

[1] Department of Physics, Technion, Haifa, Israel; soker@technion.ac.il

**Abstract:** I demonstrate the usage of planetary nebulae (PNe) to infer that a pair of jets shaped the ejecta of the core-collapse supernova (CCSN) SN 1987A. The main structure of the SN 1987A inner ejecta, the `keyhole,' comprised two low-intensity zones. The northern one has a bright rim on its front, while the southern one has an elongated nozzle. Earlier comparison of the SN 1987A `keyhole' with bubbles in the galaxy group NGC 5813 led to its identification as a jet-shaped rim-nozzle structure. Here, I present rim-nozzle asymmetry in planetary nebulae (PNe), thought to be shaped by jets, which solidify the claim that jets powered the ejecta of SN 1987A and other CCSNe. This finding for the iconic SN 1987A with its unique properties strengthens the jittering jets explosion mechanism (JJEM) of CCSNe. In a few hundred years, the CCSN 1987A will have a complicated structure with two main symmetry axes, one along the axis of the three circumstellar rings that two opposite 20,000-years pre-explosion jets shaped, and the other along the long axis of the `keyhole' that was shaped by the main (but not the only) jet pair of the exploding jets of SN 1987A in the frame of the JJEM.

## 1. Introduction

This study uses planetary nebula (PN) morphologies to strengthen the claim that jets shaped the main ejecta structure of SN 1987A, the `keyhole.' I will also mention three other core-collapse supernova (CCSN) remnants (CCSNRs). Although I cite 80 papers [1-80], this study is not a review paper but presents new results.

In an earlier study (Soker 2024c), I used the X-ray morphology of the galaxy group NGC 5813 to argue that a pair of two opposite jets shaped the `keyhole' (see Section 2) and that this pair is one out of an estimated 10-30 pairs of jets that exploded SN 1987A in the frame of the *jittering jets explosion mechanism* (JJEM; for the JJEM see, e.g., Papish & Soker 2011; Gilkis & Soker 2014; Shishkin & Soker 2021). In that paper, I used the X-ray morphologies of cooling flows in groups and clusters of galaxies to argue for shaping by jets of point-symmetric morphologies of eight CCSNRs, including SN 1987A. The significant advantage of X-ray deficient bubbles (cavities) morphologies in cooling flows is that radio observations indicate that jets from the central active galactic nucleus inflate them (e.g., Birzan et al. 2004). The disadvantages are that there is a relatively small number of cooling flows that have morphologies that are like those of CCSNRs and that the ambient medium into which the jets expand is subsonic and not expanding; CCSN ejecta expands at high supersonic velocities.

Here comes the group of bipolar PNe, i.e., those with one or more pairs of lobes. Firstly, as several PN catalogs show (e.g., Balick 1987; Chu et al. 1987; Schwarz et al. 1992; Corradi & Schwarz 1995; Manchado et al. 1996; Sahai & Trauger 1998; Hajian et al. 2007; Sahai et al. 2011; Parker et al. 2016; Parker 2022; Ritter et al. 2023; Tan et al. 2023), there are over a hundred of PNe that we can compare with CCSNR morphologies. Secondly, although not in all PNe there are direct indications for jets because the jets are long gone, there is a majority community support for the understanding that jets shaped all these bipolar PNe (e.g., Morris1987; Soker 1990; Sahai & Trauger1998; Sahai et al. 2000; Sahai et al. 2011; Boffin et al. 2012; Fang et al. 2015; Guerrero et al. 2021; Moraga Baez et al. 2023).

In an earlier paper (Soker 2024c), I identified the `keyhole' of SN 1987A to be a *rim-nozzle asymmetry*, in which the front of one bubble (cavity; lobe) in a pair is closed by a rim, which is a dense cap in three dimensions, or a partial shell, projected as an arc on the plane of the sky, while the opposite bubble (cavity; lobe) has a nozzle on its front, namely, a low-intensity opening. I present the rim-nozzle asymmetry in Section 2. I also show there the new high-quality observations by Rosu et al. (2024) that allowed me to identify the rim-nozzle asymmetry in the ejecta of SN 1987A; in some places, I will refer to SN 1987A as the CCSNR 1987A. In Section 3, I present PNe with the rim-nozzle asymmetry; in Section 4, I study three other CCSNRs with the rim-nozzle asymmetry. I postpone all discussions of the rim-nozzle asymmetry to Section 5, where I summarize the main result of this study and challenge the supporters of the neutrino-driven explosion mechanism to explain the point-symmetric CCNRs and the rim-nozzle CCSNRs without exploding jets, particularly CCSNR 1987A.

## 2. The rim-nozzle asymmetry

The high-quality HST observations that Rosu et al. (2024) published recently led me to identify the `keyhole' structure of the SN 1987A ejecta as having rim-nozzle asymmetry (Soker 2024c). In Figure 1, I present images that Rosu et al. (2024; for the keyhole see also Matsuura et al. 2024) took 12,980 days after the explosion in nine filters. My marks on the figure identify the rim-nozzle asymmetry. The structure of the [Fe I] 1.443 $\mu$m emission, as Larsson et al. (2023) present in their Figures 3 and 4, is correlated with the keyhole. The two high-intensity [FeI] zones coincide with the two bubbles of the keyhole. This strengthens the case that the axis of the keyhole is a major component of the ejecta morphology.

In principle, the magneto-rotational supernova explosion mechanism, where there is a rapid pre-collapse rotating core that leads to the launching of jets along one axis (fixed axis jets), can account for some properties of CCSNRs, e.g., nucleosynthesis (e.g., Reichert et al. 2023). However, in rare cases, the pre-collapse core has rapid rotation; therefore, this scenario can account for a fraction of CCSNe. According to the JJEM, a black hole might be formed because the fixed-axis jets are less efficient in exploding the stellar material in and near the equatorial plane in the fixed-axis jet explosion (e.g., Soker 2023a). The magneto-rotational supernova explosion can explain only one symmetry axis (e.g., Reichert et al. 2023), but not two or more, i.e., a point symmetric morphology. It seems that SN 1987A has a point-symmetric morphology. In addition, all the explosion energy is along the one symmetry axis in the magneto-rotational supernova explosion. In these cases, the energy of the two opposite jets along that axis is about the explosion energy. In the JJEM, on the other hand, there are up to tens of pairs of jets, such that the energy along the main axis of a CCSNR can be a small fraction of the total explosion energy, as deduced for many CCSNRs (Grichener & Soker 2017).

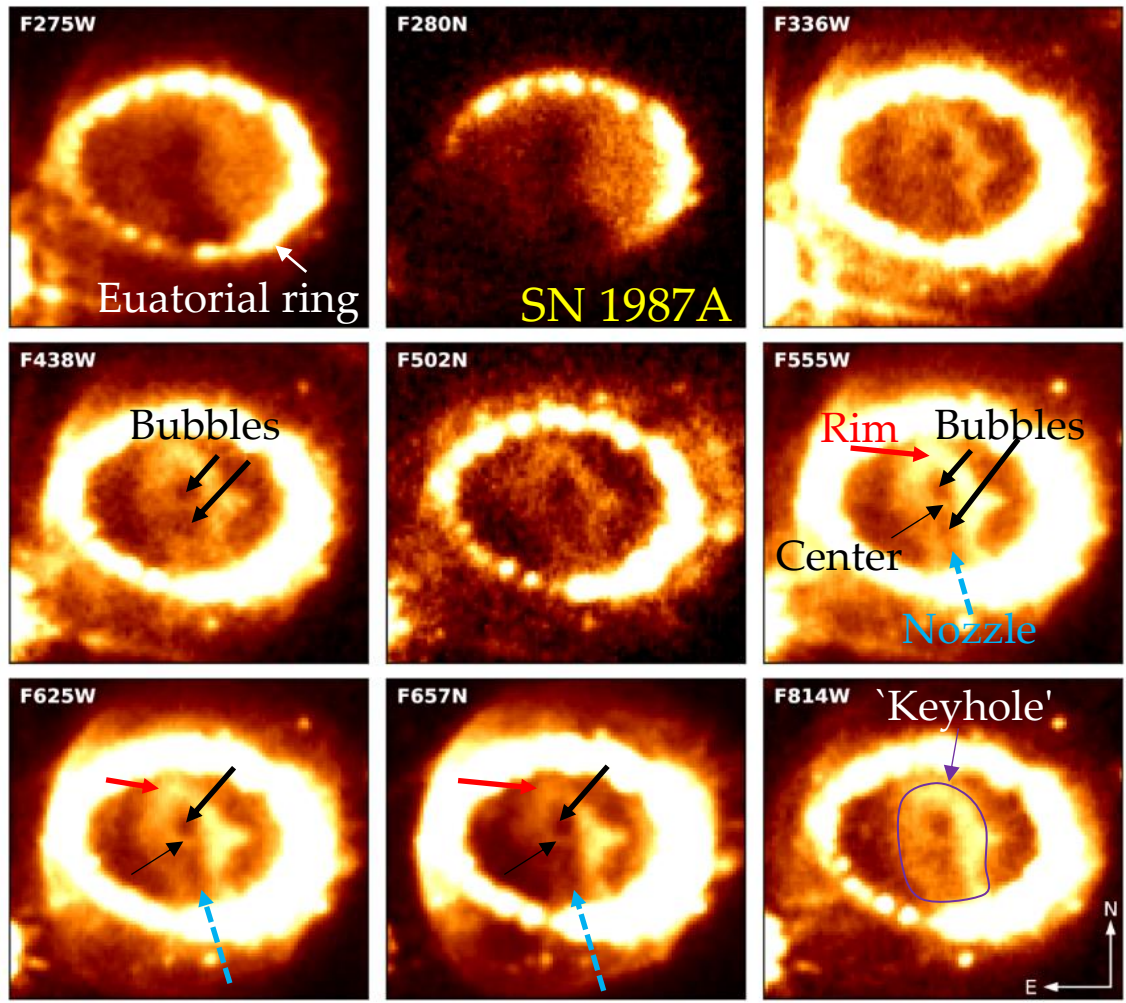

**Figure 1.** HST/WFC3 images of SN 1987A in nine filters and at 35.5 years post-explosion adapted from Rosu et al. (2024), with the marks identifying the bubbles, nozzle, and rim, from Soker (2024c). The bright structure with the low-intensity inner zones inside the purple line in the lower-right panel is the `keyhole.' The north bubble with its front rim (red arrow) and the south bubble with its nozzle (dashed-light-blue arrow) form the rim-nozzle asymmetry.

In an earlier paper (Soker 2024b) with a later update (Soker 2024c), I tentatively identified the inner ejecta of SN 1987A to have a point-symmetric structure composed of the `keyhole' and three pairs of opposite clumps. Identifying several point-symmetric CCSNRs in 2023-2024 is a significant breakthrough in establishing the JJEM as the primary, and probably the sole, explosion mechanism of CCSNe. In a recent paper, Matsuura et al. (2024) emphasize other morphological features in their high-quality JWST images of SN 1987A. In particular, the two `crescents' they identify, which are two opposite arcs outside the `keyhole' and sharing the same symmetrical axis, strengthen the claim for shaping by jets. The crescents (arcs) are another indication of shaping by jets. Essentially, the two crescents are the projection on the sky of a barrel-shaped structure. Earlier studies supported the JJEM of CCSNe by comparing several barrel-shaped PNe, known to be shaped by jets, to barrel-shaped CCSNRs (e.g., CCSNR G292.0+1.8 and SNR G309.2-00.6 studied by Soker 2023b; RCW 103 studied by Bear et al. 2017). In section 4, I present SNR G309.2-00.6, which, like SN 1987A, has both the rim-nozzle asymmetry and two crescents (arcs).

Figure 2 shows that the galaxy group NGC 5813 has several bubbles. I refer to the inner pair of bubbles, which possesses a rim-nozzle asymmetry, as the marks in Figure 2 indicate. Radio emission fills the bubbles of the galaxy group NGC 5813 (Randall et al. 2011), implying that pairs of jets inflated them. The galaxy group HCG 62 also possesses a rim-nozzle asymmetry; the radio image of HCG 62 shows leakage from the nozzle (images from, e.g., Gitti et al. 2010). These groups of galaxies imply without doubt that pairs of jets shaped the rim-nozzle asymmetry. The jet-shaped rim-nozzle asymmetry in cooling flows strongly suggests that a pair (or more) of jets also shaped the `keyhole' of SN 1987A (Soker 2024c). In Section 3, I present PNe with rim-nozzle asymmetry to solidify this claim further.

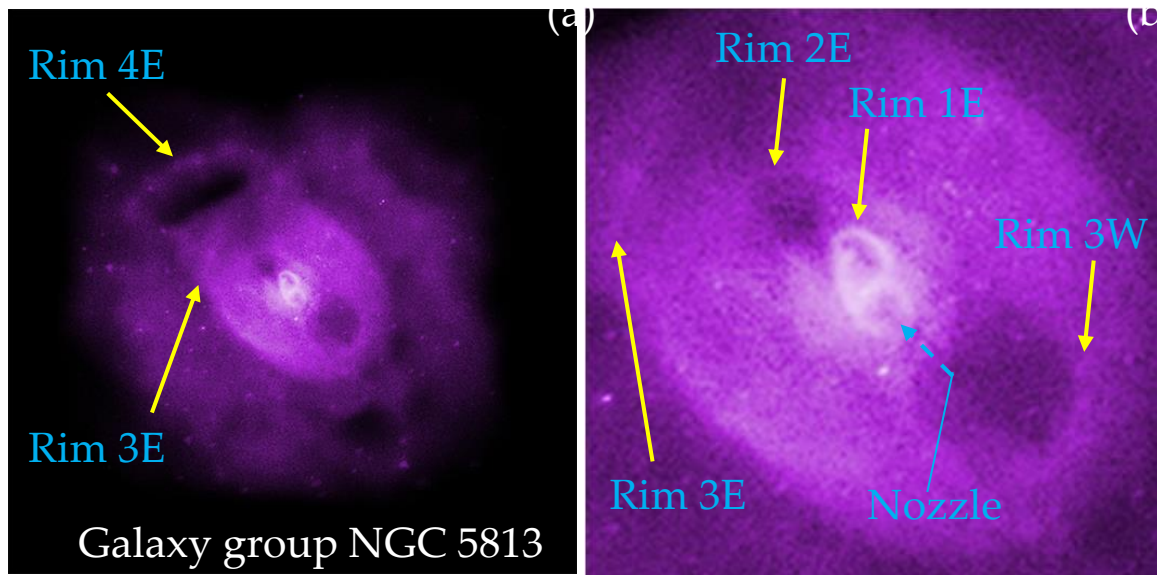

**Figure 2.** (a) Chandra 80 kpc by 80 kpc X-ray smoothed 0.3–3 keV image of the cooling flow galaxy group NGC 5813 (adapted from the Chandra site and based on Randall et al. 2015; credit: NASA/CXC/SAO/S, Randall). (b): The inner 27 kpc by 27 kpc of the left panel. Marks on both panels are from Soker (2024c). Note the *rim-nozzle asymmetry* of the bright inner structure: the northern bubble has rim 1E on its front, while the southern bubble has a nozzle (low-intensity zone) on its front.

Fransson et al. (2024) identified a zone at the center of the ejecta of SN 1987A that they argue is excited by a central neutron star or its pulsar wind nebula (for early less conclusive analysis, see Greco 2021 and Dohi et al. 2023). This region is unresolved, and it is much smaller than the keyhole. Therefore, if the pulsar wind nebula exists, it cannot explain the structure of the keyhole.

I note that while shocked jets' material emits radio waves in clusters and groups of galaxies, the radio emission in the remnant of SN 1987A results from the collision of the ejecta with the circumstellar material (CSM), mainly the equatorial ring (e.g., Zanardo et al. 2013 and Petruk et al. 2023). Petruk et al. (2023) calculate the ratio polarization that results from the ejecta-ring interaction. Future studies should include jets in the collision that gives rise to radio emission, the gas-excitation distribution (e.g., Larsson et al. 2023), the velocity field (e.g., Jones et al. 2023, Larsson et al. 2023), the interaction with the equatorial ring (e.g., Ravi et al. 2024), and other observations to better explore the role of jets in energizing and shaping the ejecta of SN 1987A. In the present study, I focus on morphologies.

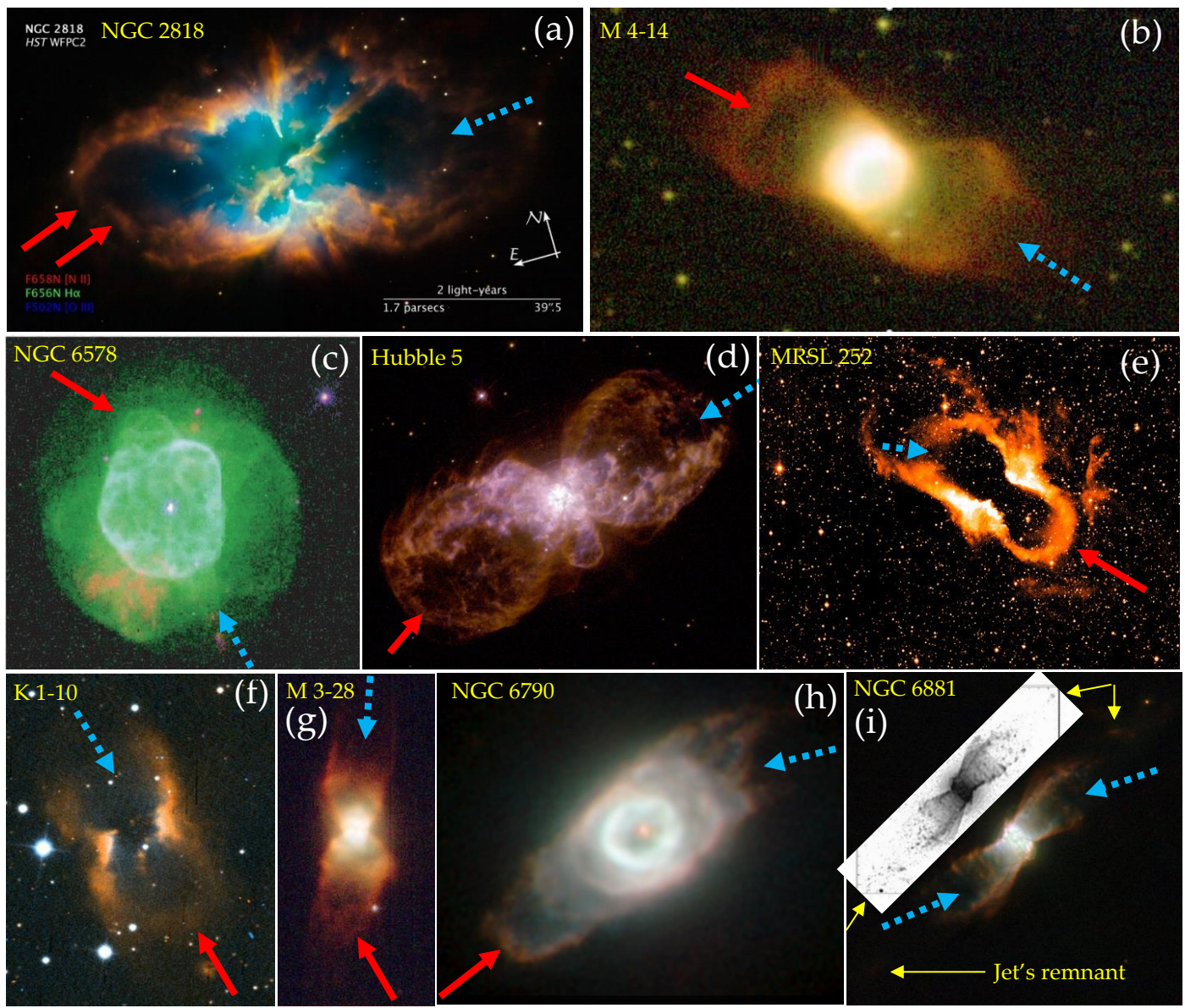

**Figure 3.** A collection of eight PNe that possess rim-nozzle asymmetry (panels a-h). The solid-red arrows point at the rims, and the dashed-light-blue arrows at the nozzles. NGC 6881 (panel i) has two nozzles with the remnants of the jets (pointed at with thin yellow arrows but can be seen only in the inset; see also Ramos-Larios et al. 2008) that probably shaped the nozzles. The inset is an image from Kwok & Su (2005), in which low-ionization gas better shows the remnants of the jets.
Sources and credits: NGC 2818: NASA, ESA, and the Hubble Heritage Team; M 4-14 + M 3-28: Manchado et al. (1996); NGC 6578: Palen et al. (2002); Hubble 5: Bruce Balick, Vincent Icke, Garrelt Mellema and NASA/ESA; MRSL 252: Corradi et al. (1997); K1-10: Schwartz et al. (1992); NGC 6790: Hubble site; NGC 6881: HST NASA/ESA, with inset from Kwok & Su (2005).

### 3. Rim-nozzle asymmetry in planetary nebulae

In Figure 3, I present eight PNe with rim-nozzle asymmetry. I mark the rims with red arrows and the nozzle with dashed-light-blue arrows. The PN NGC 6881 (Figure 3i) has two nozzles rather than a rim-nozzle asymmetry. I present this PN to show that jets shaped the nozzles (thin yellow arrows point at the remnants of the jets in the main image and the inset; the jets are visible only in the inset which shows the low-ionization gas of NGC 6881). NGC 6881 has a complicated nebula beyond the one shown in Figure 3i, as studied by Ramos-Larios et al. (2008), who suggest shaping by jets. Many other PNe have two unequal opposite lobes shaped by jets. However, their complicated and messy structure and inclination do not allow for identifying a rim-nozzle asymmetry, e.g., NGC 5189 (see images by Danehkar et al. 2018). The same holds for some CCSNRs. The CCSNR DEM 34a has lobes (see figures by Meaburn 1987). However, it possesses too large a departure from point symmetry to allow the identification of a rim-nozzle asymmetry, although it might have one. Another CCSNR where I cannot immediately identify a rim-nozzle asymmetry but which might be revealed under deeper analysis is the Cygnus Loop (see image by, e.g., Fesen et al. 2018).

**4. Other CCSNRs with rim-nozzle asymmetry**

This section presents three more CCSNRs with rim-nozzle asymmetry, each emphasizing another property.

*4.1. A multirim CCSNR*

The upper-left panel of Figure 4 is an image of the newly identified SNR candidate G107.7-5.1 (Fesen et al. 2024). In an earlier paper (Soker 2024c), I identified a nozzle, three rims, and a possible bubble, as the marks on that panel show. Based on its highly aspherical morphology, I assumed it is a CCSNR. The upper-right panel of Figure 4 presents the radio emission (in blue) from the jets in the cooling flow galaxy Hercules A that is at the center of a cluster (for more details and images, see Timmerman et al. 2022). The radio emission clearly shows that multiple jets to the west (right side) shaped a multi-rim structure (yellow arrows mark two rims). Hercules A shows that jets can form a multi-rim structure, as observed in SNR candidate G107.7-5.1. The lower panel of Figure 4 presents the PN KjPN 8 adapted from Lopez et al. (2000), who argued that pairs of jets along the two axes marked by black lines shaped the lobes. SNR candidate G107.7-5.1 and PN KjPN 8 have several rims on one side. I take the multiple-rim lobe in both the CCSNR and the PN to imply that there were several jet-launching episodes along (almost) the same direction; this is supported by Hercules A. In the case of CCSNe, the JJEM allows for some jet-launching episodes to be at only slight angles to each other. This comparison further strengthens the case of jet-shaping of SNRs.

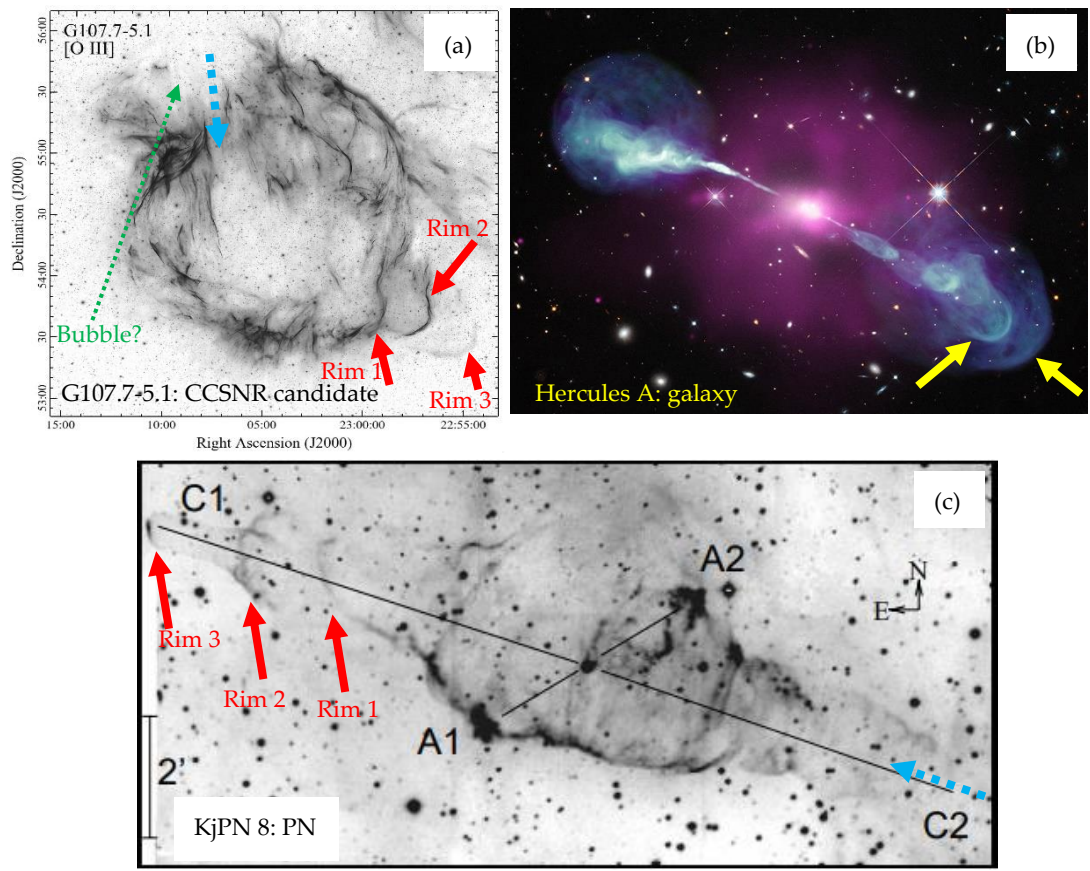

**Figure 4.** (a) An image of a new CCSNR candidate G107.7-5.1 adapted from Fesen et al. (2024). Marks of the nozzle (dashed light-blue arrow), bubble (dashed-green arrow), and rims are as in Soker (2024c). (b) Image of the galaxy Hercules A in radio (blue), X-ray (purple), and optical (white). The radio shows the jets that inflate multiple rims; two are marked by the yellow arrows (Credit: X-ray: NASA/CXC/SAO, Optical: NASA/STScI, Radio: NSF/NRAO/VLA). (c) Hα image of the PN KjPn 8 adapted from Lopez et al. (2000). I added the marks of the nozzle (dashed-light-blue arrow) and the three rims.

*4.2. A point-symmetric CCSNR*

In Soker (2023b), I argued that the point-symmetric structure of the Vela CCSNR results from the JJEM, where the last several pairs of jets left their imprints on the Vela CCSNR. In Soker (2024c), I examined new observations by Mayer et al. (2023) and identified another pair of clumps, namely H-H2 (Figure 5), solidifying the JJEM for the CCSN that formed Vela. Other models for forming jet-like structures that involve the interaction of the CCSN ejecta with an ambient gas, like the one that Velazquez et al. (2023) simulated, and instabilities in the interaction of the ejecta with an ambient gas (e.g., Chiotellis et al. 2024), cannot create such point-symmetric morphologies. The magneto-rotational supernova explosion mechanism (that has a fixed jet axis) cannot explain point symmetric CCSNRs.

Clumps E and F are significantly larger distances from the center than their counter-clumps D and J, respectively, and clump B has no counter-clump. I tentatively suggest that there are nozzles in the places I mark in Figure 5b. Deeper X-ray observations of CCSNR Vela will determine where there are indeed nozzles in the places I mark.

I argue that the possible identification of nozzles further solidifies the jet-driven explosion for the CCSN that formed Vela. The JJEM is the only theoretical model compatible with this claim in a point-symmetric CCSNR.

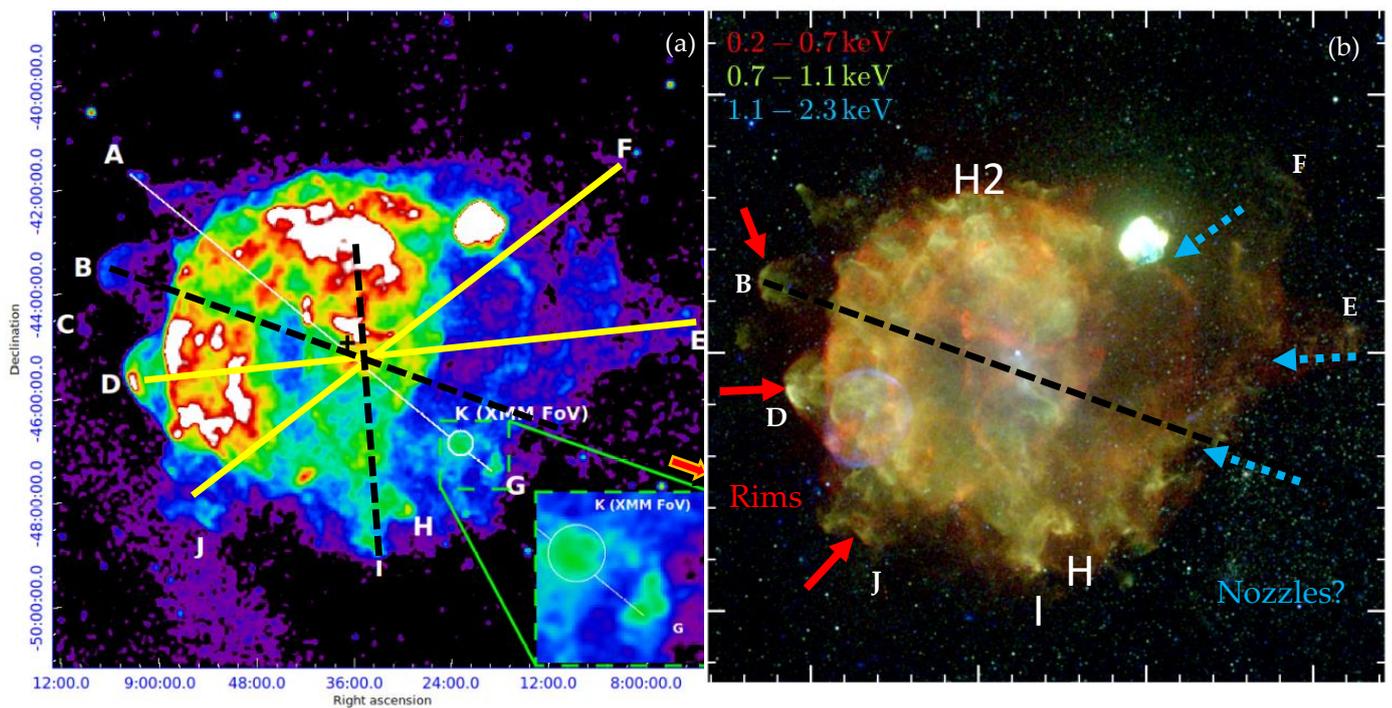

**Figure 5.** (a) ROSAT X-ray image of Vela CCSNR based on Figure 1 from Sapienza et al. (2021), who marked the clumps and white line; the original X-ray image and identification of clumps A-F are from Aschenbach et al. (1995). The thick-yellow DE-line, thick-yellow FJ-line, and the two dashed-black lines are additions from Soker (2023b). (b) Color-coded (inset) exposure-corrected eROSITA X-ray image of Vela adapted from Mayer et al. (2023). The new eROSITA X-ray image allowed me (Soker 2024c) to identify clumps H and H2 at equal distances from the center (solid-red line), solidifying the classification of a point-symmetric CCSNR. Here, I added the marks of three possible rim-nozzle asymmetry pairs.

*4.3. A Crescents (arcs) that are the projection of a barrel-shaped structure*

Some planetary nebulae have two opposite arcs (`crescents'), one at each side of the symmetry axis, with their common symmetry axis along the long PN axis. These arcs are the projections on the sky of the sides of a barrel-shaped structure. Past studies (e.g., Bear et al. 2017, Soker 2022) analyzed CCSNRs and PNe with two opposite arcs parallel to the symmetry axis of the respective nebula. Simulations (e.g., Akashi et al. 2018) show that jets can form barrel-shaped nebulae and remnants. Some CCSNRs have barrel-shaped morphology (see section 1 for a partial list) similar to the morphologies of several PNe; as jets shape PNe with arcs, this similarity indicates CCSNR shaping by jets (e.g., Soker 2023b). I here emphasize the two CCSNRs with rim-nozzle asymmetry and arcs (crescents).

Figure 6 presents the radio morphology of SNR G309.2–00.6 from Gaensler et al. (1998). I added my identification of the rim, nozzle, and the two opposite arcs. I also marked the symmetry axis with a solid yellow line, the same symmetry axis of the arcs and the rim-nozzle structure. This symmetry axis, parallel to the two pairs of arcs on the sides, is based on barrel-shaped planetary nebulae and simulations of jet-shaping (e.g., Akashi et al. 2018). Note that Gaensler et al. (1998) mention the existence of a jet inside what I identify as a nozzle. I also note that Chiotellis et al. (2021) consider the ears (the rim and nozzle) to be in the equatorial plane rather than the polar directions as I take them to be.

Matsuura et al. (2024) identified two crescents in their JWST observations of SN 1987A. These are the same structural components as the two opposite arcs that the earlier studies mentioned above identified in several other CCSNRs and PNe. I present their identification of structural features in SN 1987A in Figure 7. The two opposite arcs (from PNe with arcs) and the rim-nozzle asymmetry (from cooling flows and PNe) strongly suggest that energetic jets shape these CCSNRs and that the jets played a crucial role in the explosion process, as expected in the JJEM.

Since Matsuura et al. (2024) mention the two outer and equatorial rings, I comment on the formation of these three rings. In Soker (2022), I compared the three-ring structure of SN 1987A to the PN MyCn 18 and the clumpy equatorial ring to the Necklace nebula (IPHASX J194359.5+170901); both PNe have remnants of jets. I concluded that jets shaped the three CSM rings of SN 1987A. I now claim that jets also shaped the ejecta of SN 1987A but along a different main axis.

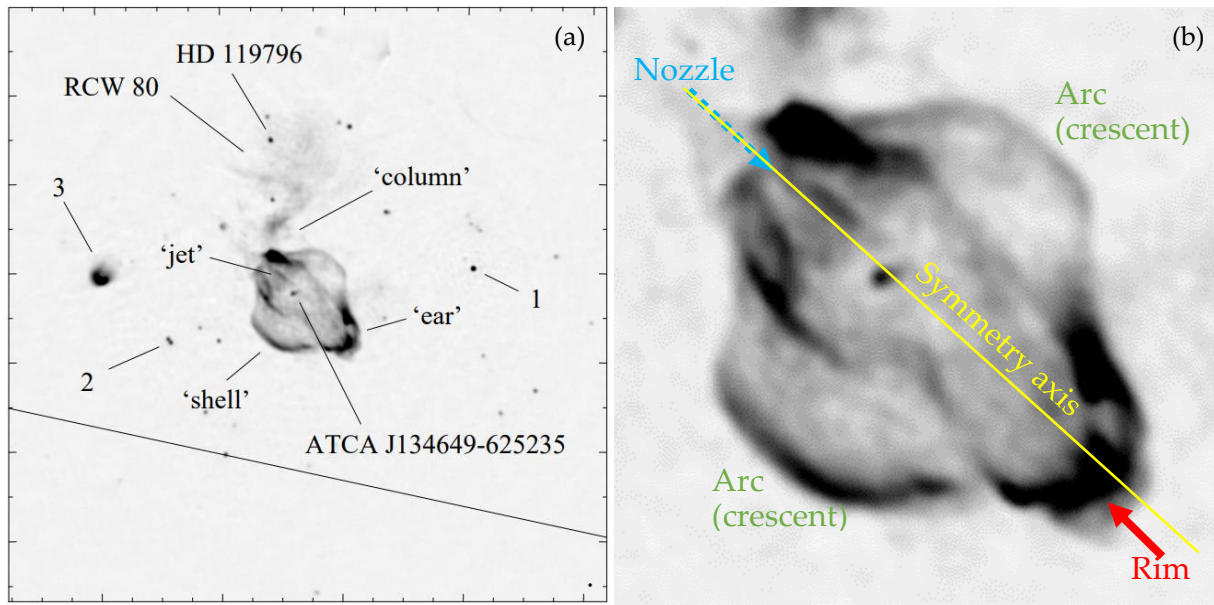

**Figure 6.** (a) A radio image of SNR G309.2–00.6 from Gaensler et al. (1998), with marks from the original image. Note that they identified a jet inside what I identify as a nozzle! (b) Focusing on the inner area of the left panel with my identification of the rim-nozzle asymmetry, the two arcs (`crescents'), and the symmetry axis of the arcs and the rim-nozzle structure (yellow line).

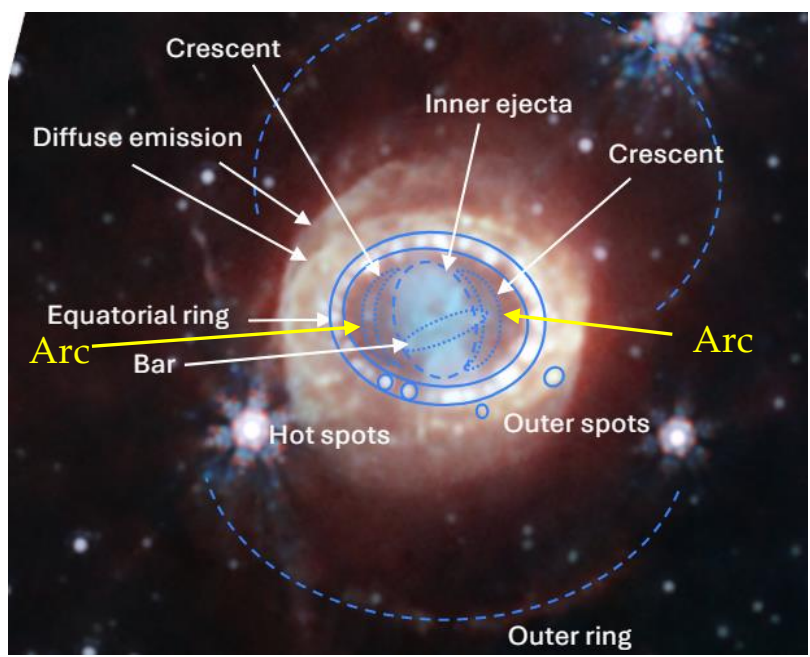

**Figure 7.** A JWST image of SN 1987a with marks from Matsuura et al. (2024). I added the marks of arcs, as I identified their termed `crescents' with arcs as used in earlier studies of CCSNRs.

## 5. Discussion and Summary

I showed that several PNe (Figure 3) possess rim-nozzle asymmetry and that these morphological structures share some properties with the rim-nozzle asymmetry of the inner SN 1997A ejecta, i.e., the `keyhole' (Figure 1). The identification of the `keyhole' as a jet-shaped structure was first made (Soker 2024c) by identifying similarities to the X-ray morphology of the cooling flow galaxy group NGC 5813 (Figure 2), which is observed to be shaped by jets.

SN 1987A is a key CCSN because we know the progenitor; it was a blue supergiant rather than a red supergiant. A blue supergiant implies no massive pre-explosion wind. We know there was no compact circumstellar material, only the equatorial ring that had not yet influenced the `keyhole' structure. No pulsar wind nebula shaped the ejecta. The `keyhole' is an outcome of the explosion process. My main claim in a previous paper (Soker 2024c) that I substantially strengthened here is that the `keyhole' is a structure of rim-nozzle asymmetry. In cooling flows (Soker 2024c) and PNe I presented in this study, jets shape rim-nozzle structures. I confidently conclude that jets shaped the `keyhole' of CCSNR 1987A and were energetic enough to be part of many pairs of jets that exploded SN 1987A, namely, the JJEM.

I further strengthened the case of the JJEM by studying three more CCSNRs with rim-nozzle asymmetry. I compared the multi-rim CCSNR candidate G107.7-5.1 with the jet-shaped multi-rim cooling flow Hercules A and PN KjPn 8 (Figure 4). Most likely, three jet-launching episodes, compatible with the JJEM, shaped the three rims of CCSNR candidate G107.7-5.1. I tentatively identified three rim-nozzle asymmetry pairs in the point-symmetric CCSNR Vela (Figure 5b). That a fraction of CCSNRs have point-symmetric morphology is a fundamental expectation of the JJEM. Identifying rim-nozzle pairs, even if tentative, strengthens the claim of jet-shaping of the Vela SNR. The SNR G309.2–00.6 possesses a rim-nozzle asymmetry and two opposite arcs sharing the same symmetry axis (Figure 6). Both these morphological structures support shaping by jets, as I discuss in section 4.3. The two opposite arcs are the projection on the plane of the sky of a barrel-shaped structure. In a recent study, Matsuura et al. (2024) identified two such arcs, which they termed `crescents,' in the ejecta of SN 1987A. The two arcs further strengthen the case of jet-shaping of CCSNR 1987A, adding to the rim-nozzle asymmetry that was studied here following Soker (2024c) and to the point-symmetry of SN 1987A (Soker 2024b, 2024c).

The morphological indications for shaping CCSNR ejecta by jets, and in many cases by multi-jet-pairs as expected in the JJEM, and where the jets are energetic enough to explode the star, add to other supporting arguments of the JJEM (for reviews, see Soker 2022, 2024b). I bring one such argument here. The lightcurves of many CCSNe require extra energy in addition to the initial explosion energy and the radioactive decay of nickel. Interaction with a circumstellar medium does not always account for that. Therefore, researchers commonly suggest a central magnetar, i.e., a rapidly spinning highly-magnetized neutron star (e.g., Rodriguez et al. 2024 for a very recent study). However, energetic jets that explode the star likely accompany the formation of energetic magnetars (e.g., Soker 2016, Soker & Gilkis 2017). Late jets that add to the magnetar power can follow the exploding jets (Soker 2022). Namely, the presence of an energetic magnetar also implies exploding jets. The JJEM accounts for jets in all CCSNe, not only a subclass of CCSNe.

The cause of a rim-nozzle asymmetry might be that on the side of the nozzle, the jet is more energetic and/or narrower (i.e., has a larger momentum flux on the axis) than the counter-jet and/or that on that side, the mass of the ambient medium is lower than on the opposite side (the side of the rim). Akashi & Soker (2021) simulated different cases and found that the jet might form a rim early, but later, it breaks out to create a nozzle (as the lower panels of their figure 6 show). Here, I assume that the stronger jet can open the nozzle when turned off, but not the weaker one. In Soker (2024a), I discussed the rim-nozzle asymmetry in the JJEM. In the JJEM, a typical jet-launching episode lasts for $\tau_{\rm jet} \simeq 0.01-0.1 \, {\rm s}$ a timescale that is not much longer, or even shorter, than the relaxation time of the accreted material to a thin accretion disk. The accreted material has no time to relax to a thin accretion disk; the two opposite sides of the thick accretion disk will likely differ in the exact structure and magnetic field properties. Therefore, the disk's sides likely launch two opposite jets that differ in power. In cooling flows, the intracluster medium is likely to be non-homogeneous. Thus, the inflated bubbles might differ even if the two opposite jets are identical.

What is the cause of the rim-nozzle asymmetry in PNe? I speculate that in PNe, one reason for unequal opposite jets might be the feeding of the star that launches the jets: a main sequence star or a white dwarf. Consider an asymptotic giant branch (AGB) star that feeds its jet-launching companion via a Roche lobe overflow (RLOF). This process might sometimes not be symmetric on both sides of the equatorial plane because of the large and fast convective cells in the AGB envelope. Convection motion is stochastic, and the convective cells are giant in AGB stars. These properties imply that a giant convective cell may rise to the surface on one side of the equatorial plane, pushing more gas toward the companion from that side. No cell is on the other side of the equatorial plane at that specific time. In other words, a sizeable convective cell moving outward on one side of the equatorial plane might cause a protrusion of the AGB envelope. The protruding side might cause extra mass flow, leading to an accretion disk's unequal sides that launch jets of unequal power. This phase might last for an AGB dynamic timescale of about a year or even longer. This timescale might be sufficient to set a pronounced asymmetry if the jets are very energetic during that time, as in the case of an intermediate luminosity optical transient (ILOT) event. ILOTs might occur as part of the formation process of some bipolar PNe (e.g., Soker & Kashi 2012). This speculation of forming rim-nozzle asymmetry in PNe requires further study.

Although **I consider this study to be another step in establishing the JJEM as the primary explosion mechanism of CCSNe**, it is mandatory to mention another alternative/competing theoretical explosion mechanism, namely, the delayed neutrino explosion mechanism (e.g., Wongwathanarat et al. 2013; Burrows & Vartanyan 2021; Orlando et al. 2021; Fryer et al. 2022, 2023; Mezzacappa 2023; Andresen et al. 2024; Boccioli & Roberti 2024; Burrows et al. 2024; Janka & Kresse 2024; Muller et al. 2024; van Baal et al. 2024; and, e.g., Utrobin et al. 2021 for SN 1987A). The delayed neutrino explosion mechanism and the JJEM share some properties, like the requirement of vigorous convection in the pre-collapse core. Also, neutrino heating plays a role in the JJEM in boosting the explosion, but not the primary role. The delayed neutrino mechanism suffers from some consistency problems, e.g., the different groups do not agree on some even qualitative results. For example, various research groups do not agree on the stellar models that explode or fail to explode. Note that in the JJEM, there are no failed CCSNe (see review by Soker 2024b). In the last year, the most severe challenge to the delayed neutrino explosion mechanism has been explaining the formation of point-symmetric CCSNRs, which studies recently revealed (Soker 2024b, 2024c). The rim-nozzle asymmetry I studied in the present work is another morphological feature for which the delayed neutrino explosion mechanism should account. *I challenge the supporters of the neutrino-driven explosion mechanism to explain the point-symmetric CCNRs and the rim-nozzle CCSNRs.*

In section 4.3, I mentioned that the three circumstellar rings of SN 1987A share some properties with the PNe MyCn 18 and the Necklace. I took it to imply that jets shaped the three rings. The symmetry axis of the three rings is highly inclined to that of the `keyhole.' I predict that in several hundred years (*In the year 2525*), after most of the ejecta interact with the outer rings, there will be two main symmetry axes of CCSNR 1987A, one determined by the pre-explosion jets that shaped the three rings and one by the main jets that shaped the `keyhole.'

**Funding:** This research was funded by the Pazy Foundation.

**Acknowledgments:** I thank the four referees for very useful comments and for pointing out missing references and relevant objects. I used the Planetary Nebula Image Catalogue (PNIC) compiled by Bruce Balick: https://faculty.washington.edu/balick/PNIC/ .

**Conflicts of Interest:** The author declares no conflict of interest.